\documentclass{svproc}
\usepackage{url}

\usepackage{amsfonts}
\usepackage{amsmath}
\usepackage{amssymb}
\usepackage{mathrsfs}
\usepackage{latexsym}
\usepackage{multicol}
\usepackage{fancybox}
\usepackage{color}
\usepackage{hyperref}
\usepackage{xcolor}
\usepackage{graphicx}
\usepackage{caption}
\usepackage{enumerate}
\usepackage{upgreek}
\usepackage[overload]{empheq}
\def\be{\begin{equation}}
\def\ee{\end{equation}}

\def\d'{``}

\begin{document}
\mainmatter              % start of a contribution
\title{Non rectification of heat in graded Si-Ge alloys}
\titlerunning{Non rectification of heat in graded Si-Ge alloys}  % abbreviated title (for running head)
\author{S. Carillo\inst{1} \and M.G. Naso\inst{2} \and E. Vuk \inst{2} \and F. Zullo \inst{2}}
%
%\authorrunning{Ivar Ekeland et al.} % abbreviated author list (for running head)
%Craig Chambers, Kim B. Bruce, and Elisa Bertino}
%
\institute{Dipartimento di Scienze di Base e Applicate
per l'Ingegneria,\\  Universit\`a di Roma ``\textsc{La Sapienza}'',
Via Antonio Scarpa 16, 00161 Rome, Italy,\\ \&
    \\	I.N.F.N. - Sezione Roma1,
Gr. IV - M.M.N.L.P.,  Rome, Italy \\ 
\email{sandra.carillo@uniroma1.it},%\\ WWW home page:
%\texttt{http://users/\homedir iekeland/web/welcome.html}
\and
DICATAM, Universit\`a degli Studi di Brescia, Via Valotti 9, Brescia, Italy\\
\email{mariagrazia.naso@unibs.it}, \email{elena.vuk@unibs.it}, \email{federico.zullo@unibs.it}}

\maketitle              % typeset the title of the contribution

\begin{abstract}
We investigate the possibility to obtain a thermal diode with functionally graded Si-Ge alloys. A wire with variable section is considered. After the introduction of a formula giving the thermal conductivity of the wire as a function of the species content and of the diameter of the wire, numerical and analytical results are presented supporting the impracticability to get a thermal diode with the characteristics here considered.{However, the present study opens the way to further generalisations amenable to give applicative promising results. }
\end{abstract}


\section{Introduction} \label{intro}
The thermal phenomenon that allows the heat to be transferred in a suitable direction in a given material, while the flow is impeded in the opposite direction, is called thermal rectification \cite{noi,P}. This is the analogue of the current rectification of the electronic diodes and for this reason any device showing some thermal rectifying feature is called thermal diode. An homogeneous material possessing a constant thermal conductivity is known to not possess any rectifying  property \cite{GS}: the heat flows, under the same thermal gradient applied, equally in all the directions. It follows that, if a material possesses a rectifying effect, {{then}} the thermal conductivity $\lambda$ {{is }} a non-homogeneous function of the temperature. With non-homogeneous we mean that the thermal conductivity  also  depends  on the space variable $\textbf{x}$. Despite to be necessary, this condition is however far from being sufficient: indeed, it has been shown \cite{GS} that, if the thermal conductivity is a separable function, i.e. if there exist two functions $f$ and $g$ such that $\lambda(T,\textbf{x})=f(T)g(\textbf{x})$, then no rectifying effect {{can}} be observed in the material.

A practicable process to obtain non-homogeneous values of macroscopic properties, such as the thermal conductivity, is the manufacturing of functionally graded materials, i.e. materials with a specific gradation in the composition in order to achieve particular performances or functions \cite{MDSK,NS}. One of the most common example of graded materials are materials made of binary alloys $A_cB_{1-c}$, where $A$ and $B$ are two different atomic species and $c$ is the content of the specie $A$ (so that $c\in (0,1)$).  In this work we present a detailed analysis of the possibility to get a thermal diode with graded Si-Ge alloys. The investigation is mainly based on the results given in \cite{noi}, where a systematic approach has been introduced to find the optimal gradation of the species in order to maximize the efficiency of the fin.  Our main assumptions are the following {{ones}}: 
%the device is a wire of variable diameter $D$, hence the problem can be considered one-dimensional; the Fourier law describes with enough accuracy the distribution of temperature along the device; the anharmonic, alloy and boundary scattering of the phonons all give contributions to the value of the thermal conductivity; the concentration $c$ and the diameter $D$ are unknown variables and must be determined in order to optimize the rectifying effect.
{{\begin{itemize}
\item the device is a wire of variable diameter $D$, hence the problem can be considered one-dimensional;
\item the Fourier law describes with enough accuracy the distribution of temperature along the device;
\item the anharmonic, alloy and boundary scattering of the phonons all give contributions to the value of the thermal conductivity; 
\item the concentration $c$ and the diameter $D$ are unknown variables and must be determined in order to optimize the rectifying effect.
\end{itemize}}}

The paper is organized as follows: in Section~\ref{sec2} we introduce the main equations and the rectification coefficient $R$. Also, for the sake of clearness and completeness we briefly describe the approach given in \cite{noi} to maximize this coefficient. In Section~\ref{sec3} we apply the formulae to a Si-Ge graded wire with variable diameter. An analytical formula for the thermal conductivity, mainly based on \cite{WM} (see also \cite{CD,K,Kle}), is presented and discusses. In Section~\ref{sec4} we apply the results given in Section~\ref{sec2} to the formula for the thermal conductivity obtained. Finally, in the conclusions,  the main aspects of this work {{are emphasized}}, {{under}} a constructive point of view. {{Indeed, even}} if it is true that our analysis seems to preclude the possibility to get significant values of the rectification coefficient, different perspectives 
{{which}} may  {{ indicate the methodology to achieve applicable}} results.

\section{The rectification coefficient and its maximization}\label{sec2}
This section is mostly based on a work of Peyrard \cite{P} and on a previous work of some of the co-authors \cite{noi} and introduces the main findings described in Section~\ref{sec3}. It has been shown in \cite{P} that it is possible to get a rectifying effect from a device composed of two materials with at least one of them with a temperature dependent thermal conductivity. When the temperature range considered is large, almost every metallic or semiconductor material has a temperature dependent thermal conductivity. Further{{more}}, for graded materials, the thermal conductivity may be also a function of the gradation in composition: for example, for binary alloys $A_{c}B_{1-c}$, $\lambda$  is a function of $T$ and of the species content $c$ (see e.g. \cite{SC}). If the species content $c$ is variable inside the material, i.e. $c=c(\textbf{x})$, then the thermal conductivity becomes dependent on $T$ and $\textbf{x}$, $\lambda=\lambda(T,c(\textbf{x}))$. In \cite{noi} a systematic way to choose the spatial distribution of the composition $c(\textbf{x})$ and the geometry of the device presenting the more interesting rectification performances has been introduced. Here we will recall the main findings for completeness and easy of readability of Section~\ref{sec3}.
 By assuming that the Fourier law holds, the steady state distribution of the temperature is described by the solution of the following equation:
\begin{equation}\label{Fourier}
\frac{dT}{dx}=-\frac{q}{\lambda(T,c(x))}.
\end{equation}
In a steady state situation, like the one we are considering, the heat flux $q$ across the device is a constant (since it solves  $\nabla \cdot \textbf{q}=0$). Then, the implicit solution of equation (\ref{Fourier}) is given by
\begin{equation}\label{sol}
T(x)=T(0)-q\int_{0}^x\frac{1}{\lambda\left(T(y),\nu(y)\right)}dy,
\end{equation}
giving
\begin{equation}\label{q}
q=-\frac{T(L)-T(0)}{\displaystyle{\int_{0}^L\frac{1}{\lambda\left(T(y),\nu(y)\right)}dy}}.
\end{equation}
The efficiency of a thermal rectifier can be evaluated through the rectification coefficient, defined by the ratio of the heat flux in two opposite configurations, the ``direct'' and the ``inverse''. To fix  ideas, we take the two boundaries of the device at $x=0$ and $x=L$ at the temperatures $T_{H}$ and $T_{L}$, where $T_{H}>T_{L}$. In the direct configuration the end $x=0$ is at the temperature $T_L$ and the end $x=L$ at the higher temperature $T_H$. In the reverse configuration the boundary $x=0$ is at the higher temperature $T_H$ whereas the end at $x=L$ is at the lower temperature $T_L$. From equation (\ref{sol}), if $q_d$ and $q_r$ are respectively the direct and reverse heat fluxes, the rectification coefficient $R$ is defined as:
\begin{equation}\label{R1}
\displaystyle{R=\frac{|q_d|}{|q_r|}=\frac{\displaystyle{\int_{0}^L\frac{1}{\lambda\left(\tau_r(y),c(y)\right)}dy}}{\displaystyle{\int_{0}^L\frac{1}{\lambda\left(\tau_d(y),c(y)\right)}dy}},}
\end{equation}
where $\tau_r$ and $\tau_d$ are the solutions of the steady Fourier equation (\ref{Fourier}) in the reverse and direct configurations respectively \cite{noi}. From (\ref{R1}) it is clear that if $\lambda$ was a constant, the rectification constant would be equal to 1, i.e. the heat fluxes in the direct and inverse configurations are equal and no rectifying effect is observed.
Notice that, if $\lambda$  {{is}} constant,  {{then }}the distribution of the temperatures in the two configurations  {{is represented by}}
 the distributions of the temperatures $T_d$ and $T_r$ in the direct and reverse configurations, {{ given, respectively, by}} 
\begin{equation}\label{diag}
T_{d}=T_l+(T_h-T_l)\frac{x}{L}, \qquad T_r=T_h-(T_h-T_l)\frac{x}{L}.
\end{equation}
If we look at the plane $(x,T)$, these temperatures lie exactly on the diagonals of the rectangle with vertices in $(0,T_L)$, $(0,T_H)$, $(L,T_L)$ and $(L,T_H)$. 
{{According to}} \cite{noi},  if $\lambda$ is a regular continuous function of $T$ and $x$, at steady state the temperature profiles roughly will follow the same diagonals. Then, if the value that the thermal conductivity assumes on one of these diagonals is much greater then the value it assumes on the other diagonal, a considerable rectifying effect should be observed.  Based on this reasoning, in \cite{noi} the following methodology has been proposed to maximize the rectification coefficient (\ref{R1}):
\begin{enumerate}
\item Among the various possible geometries and distributions $c(x)$, look for the particular geometry and distribution $c_0(x)$ giving a saddle point to the function $\lambda(T,x)$ in the middle of the domain of interest.  In the plane $(x,T)$, the saddle must present a maximum on one diagonal and a minimum on the other diagonal. \label{it1}
\item  Among the various possible geometries and distributions $c(x)$, look for the particular geometry and the particular distribution $c_0(x)$ {{which maximizes}} the difference between the values that $\lambda(T,x)$ assumes on the vertices of the rectangle {{whose }}diagonals {{are }}given by (\ref{diag}). More precisely, if $T_A(x)$ denotes the values on one of the two diagonals and $T_B(x)$ the values on the other, maximize the differences $\lambda(T_A(0),0)-\lambda(T_B(0),0)$ and $\lambda(T_A(L),L)-\lambda(T_B(L),L)$.  \label{it2}
\end{enumerate}
In the next section, in order to apply the above line of reasoning to graded Si-Ge devices, we will describe the dependence of the thermal conductivity on the temperature, gradation and dimension of the section of the wire.

\section{Si-Ge alloys with variable section}\label{sec3}
The thermal conductivity of silicon germanium alloys is in general lower than the corresponding thermal conductivities of the bulk materials. From a microscopic point of view, this is due to an additional scattering mechanism of the phonons inside the material, the so-called alloy scattering. The geometry of the material may play also a role in the determination of the thermal conductivity. Indeed, if the dimensions are small enough, the scattering of the phonons with the boundaries of the material may become strong enough and contribute to a further lowering of the thermal conductivity. The properties of the Si-Ge alloys can be computed by using first-principles approaches, like the density-functional perturbation theory (see e.g. \cite{Li} for bulk materials with diffusive boundary conditions or \cite{Garg} for disordered silicon-germanium alloys). As for the case of porous silicon, we need an analytical formulation of the thermal conductivity accurate enough and simple enough to be manipulated. In \cite{WM} the authors presented  a theoretical, phenomenological formulation of the thermal conductivity of $\textrm{Si}_{1-c}\textrm{Ge}_{c}$ nanowire alloys, with the section of the wire explicitly taken into account. This formulation has been compared with experimental results: between a temperature range of $(100,400)$K, the agreement seems to be good enough in order to be utilized here \cite{CD}, \cite{K}, \cite{WM}. The thermal conductivity of a  $\textrm{Si}_{1-c}\textrm{Ge}_{c}$ nanowire is then given by \cite{WM}
\begin{equation}\label{lamsige}
\lambda=\frac{k_B^4T^3}{2\pi^2 v\hbar^3}\int_0^{\hbar w_c/k_BT}\frac{y^4e^y}{(e^y-1)^2}\frac{1}{\tau^{-1}}dy,
\end{equation}
where $k_B$ and $\hbar$ are the Boltzmann's and reduced Planck's constants,  the inverse scattering rate $\tau^{-1}$ is given, by the Mathiessen's rule, by the sum of three terms, i.e.
$$\tau^{-1}=\tau^{-1}_u+\tau^{-1}_a+\tau^{-1}_b,$$
where $\tau^{-1}_u$, $\tau^{-1}_a$ and $\tau^{-1}_b$ are respectively the contributions due to anharmonic, alloy and boundary scattering. The anharmonic contribution is described by the weighted average between the Si and Ge terms,
$
\tau^{-1}_u=(1-c)\tau^{-1}_{Si}+c\tau^{-1}_{Ge},
$
with $\tau^{-1}_{Si}$ proportional to $w^2Te^{-C_{Si}/T}$ and $\tau^{-1}_{Ge}$ proportional to $w^2Te^{-C_{Ge}/T}$ ($C_{Si}$ and $C_{Ge}$ are constants, $w$ is the frequency of the phonons). The alloy scattering term is approximated by a quadratic function of $c$, that must be zero for $c=0$ and $c=1$, giving $\tau^{-1}_a\sim c(1-c)w^4$ (see also \cite{Kle} for the proportionality to the $w^4$ term). The boundary term is taken in \cite{WM} to be equal to $\tau^{-1}_b=v/D$, where $D$ is the diameter of the wire and $v$ is the average speed of sound, given by $v^{-2}=(1-c)v_{Si}^{-2}+cv_{Ge}^{-2}$ ($v_{Si}$ and $v_{Ge}$ are the average speeds of sound in silicon and germanium). The overall cutoff frequency $w_c$ is given by $w_c=w_{cut} v/v_{Si}$, where $w_{cut}\sim 38.8$THz \cite{WM}.

Putting all together and taking the constants of proportionality from \cite{WM}, we get
\small
%\begin{equation}\label{lamsige1}
%\hspace{-5mm}\lambda=\hat{\lambda}_0T^3\beta\int_{0}^{\Theta_c/\beta T}\frac{y^4e^y}{(e^y-1)^2}\frac{dy}{(1-c)y^2T^3e^{-C_{Si}/T}+1.93cy^2T^3e^{-C_{Ge}/T}+3.41c(1-c)T^4y^4+96.6/(\beta D)}
%\end{equation}
\begin{equation}\label{lamsige1}
\hspace{-2mm}\displaystyle{\lambda=\hat{\lambda}_0T^3\beta\int_{0}^{\Theta_c/\beta T}\!\!\!\!\!\!\!\!\!\frac{y^4e^y (e^y-1)^{-2}}{(1-c)y^2T^3e^{-C_{Si}/T}+q_1cy^2T^3e^{-C_{Ge}/T}+q_2c(1-c)T^4y^4+q_3/(\beta D)}dy,}
\end{equation}
where $\hat{\lambda}_0=2.43 \textrm{W} \textrm{m}^{-1} \textrm{K}^{-1}$, $\beta=\sqrt{1521+2575c}$, $\Theta_c=1.1568\cdot 10^4\textrm{K}$, $C_{Si}=139.8\textrm{K}$ and $C_{Ge}=69.34\textrm{K}$. The three constants $q_i$ are given by $q_1=1.93$, $q_2=3.41\textrm{K}^{-1}$, $q_3=96.6 \textrm{K}^3\textrm{m}$.

\begin{figure}\centering
{\includegraphics[scale=0.8]{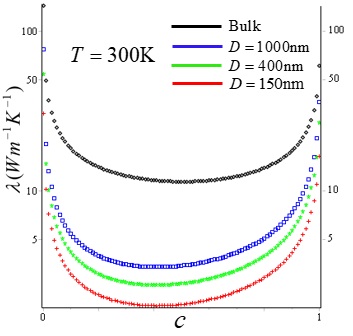}}
\caption{\small{The values of the thermal conductivity obtained from formula (\ref{lamsige1}) with $T=300\textrm{K}$ for $D=\infty$ (corresponding to a bulk alloy), $D=1000\textrm{nm}$,  $D=400\textrm{nm}$ and  $D=150\textrm{nm}$. A log scale has been used on the vertical axis.}}
\label{F7}
\end{figure}

In Figure (\ref{F7}) the plots of the value of the thermal conductivity for $T=300\textrm{K}$ and for different values of $D$ are reported. In the next subsection we apply our methodology to the thermal conductivity (\ref{lamsige1}): as we will see, in this case we find that it is not possible to obtain high values of the rectification coefficient (\ref{R1}) following our approach.

\section{Analysis of the thermal conductivity (\ref{lamsige1}) and the corresponding rectification coefficient}\label{sec4}
\begin{figure}\centering
{\includegraphics[scale=0.7]{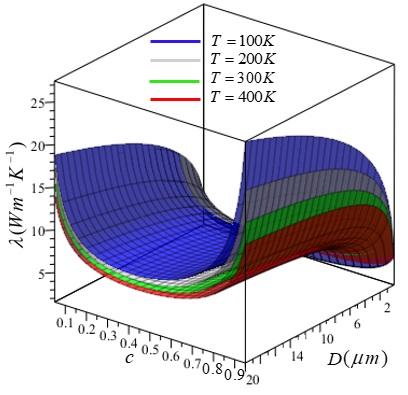}}
\caption{\small{The values of the thermal conductivity for $c\in (0.05, 0.95)$ for different values of the temperature.}}
\label{F8}
\end{figure}

In the approach explained in Section~\ref{sec2} to obtain sufficiently large values of the rectification coefficient it is crucial, for \emph{both} the steps \ref{it1}) and \ref{it2}),  to have  large values for the differences $\lambda(T_A(0),0)-\lambda(T_B(0),0)$ and $\lambda(T_A(L),L)-\lambda(T_B(L),L)$, where $T_A(x)$ defines the values of the temperatures on one of the two diagonals and $T_B$ the values on the other.
%following physical observation: if the values of the thermal conductivity on the diagonals of the rectangle with vertices $(0,T_l)$, $(0,T_h)$, $(L,T_l)$ and $(L,T_h)$ are substantially different, the direct and inverse flows will be different as well and a reasonable value of the rectification coefficient will be obtained. To quantify the words ``substantially different'' and ``a reasonable value'' we made the example showed in Figure (\ref{F2}). In both the steps \ref{it1}) and \ref{it2}) described in section (\ref{sec2}) it is crucial to have  large values for the differences $\lambda(T_A(0),0)-\lambda(T_B(0),0)$ and $\lambda(T_A(L),L)-\lambda(T_B(L),L)$, where $T_A(x)$ defines the values of the temperatures on one of the two diagonals and $T_B$ the values on the other (see also point \ref{it1}) in section (\ref{sec2})).
\begin{figure}\centering
{\includegraphics[scale=0.6]{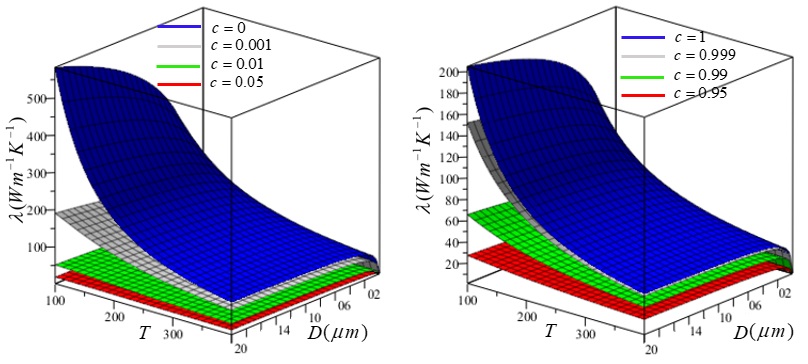}}
\caption{\small{The values of the thermal conductivity as a function of the temperature in the two dilute zones given by $c\in (0, 0.05)$ and $c\in(0.95,1)$ for some values of $c$.}}
\label{F9}
\end{figure}
It is clear however that, in order to satisfy these conditions, the function $\lambda(T,x=0)$, as a function of $T$ in the domain $(T_l, T_h)$, {{is required to be }}\emph{increasing} at least in some interval of the domain, whereas the  function $\lambda(T,x=L)$  {{is required to be }} \emph{decreasing} at least in some interval of the same domain. Since in formula (\ref{lamsige1}) we assume that both the concentration $c$ and the diameter $D$ are function of $x$, the values $x=0$ and $x=L$ pick out some values $c(0)$, $c(L)$, $D(0)$ and $D(L)$. If the function $\lambda(T,c,D)$ would be increasing, as a function of $T$, in some region of the variables $c$ and $D$, and decreasing in some other regions, then it would be possible, by tuning the dependence of $c$ and $D$ on $x$, to give a thermal conductivity presenting an interesting rectification coefficient. But, in the range of temperatures $(100,400)$\textrm{K}, the function $\lambda(T,c,D)$ {{does}} not {{satisfy}} the  requested {{properties}}. Indeed{{,}} in the region $c=(0.05,0.95)$ {{$\lambda$}} is almost flat for all values of $T$ and for all the reasonable values of $D$, i.e. from $10$nm to $2\cdot 10^4$nm. In this range of $c$, the values of $\lambda$ are bounded in the interval $(1-25) \textrm{W} \textrm{m}^{-1} \textrm{K}^{-1}$, as shown in (\ref{F8}), whereas in the full range of variation of $c$ (i.e. $c\in(0,1)$), the values of $\lambda$ can reach values as large as $500 \textrm{W} \textrm{m}^{-1} \textrm{K}^{-1}$. This suggests to look at values of $c$ giving almost pure silicon ($c\in(0,0.05)$) or almost pure germanium ($c\in(0.95,1)$). {{However,}} in these ranges the thermal conductivity is always a \emph{decreasing} function of the temperature, independently on the value the concentration assumes, apart very flat maxima for fixed and small values of $D$ (see the plots (\ref{F9}) for the dilute zones). This {{seems to }}preclude the possibility to tune the dependence of $c$ and $D$ on $x$ in such a way to get an interesting value of the rectification coefficient. The above result could explain the small values of $R$ found in literature for $\textrm{Si}_{1-c}\textrm{Ge}_c$ devices (see, e.g., \cite{CCJ}).

\section{Conclusions}
In this work we systematically analyzed the possibility to obtain a thermal diode for functionally graded Si-Ge alloys. We tried to get the particular spatial distribution of the species content $c$ and the geometry of the
wire giving reasonably high values of the rectification coefficient. This same methodology  {{is}} applied to porous silicon materials  in \cite{noi} showing the possibility to obtain a rectification coefficient equal to 3.15. The same approach applied to Si-Ge materials shows that a thermal diode with the characteristics here described has instead few chances to become
a good heat rectifier. Clearly, this negative result only implies the impracticability to get a thermal diode with the above characteristics. The model here presented, although to be physical consistent and accurate, {{can be regarded as}} a first approximation to more complex approaches and if new variables or different heat transfer laws are embedded in the model {{here considered}}, more satisfactory results could be achieved.

\section{Acknowledgments}
Work performed under the auspices of the Gruppo Nazionale per la Fisica Matematica (GNFM-INdAM).

\smallskip

\noindent M.G.N., E.V., and F.Z. would like to acknowledge the financial support of the Universit\`{a} degli Studi di Brescia. 

\noindent S.C.  acknowledges the partial financial support of Universit\`a di Roma ``\textsc{La Sapienza}''.

\smallskip

\noindent 
S.C. and F.Z. would like to acknowledge the financial support of INFN.



  \bibliographystyle{elsarticle-num}
  \bibliography{bibl}

\end{document}